\documentclass[cits]{PoS}\usepackage{amsmath,cite,color}

\title{Exotic states and their properties from large-$N_{\rm c}$
QCD}\ShortTitle{Exotic states and their properties from
large-$N_c$ QCD} 
\author{Wolfgang Lucha\\Institute for High Energy Physics, Austrian
Academy of Sciences, Nikolsdorfergasse 18, A-1050 Vienna,
Austria\\E-mail: \email{Wolfgang.Lucha@oeaw.ac.at}}
\author{\speaker{Dmitri Melikhov}\\Institute for High Energy
Physics, Austrian Academy of Sciences, Nikolsdorfergasse 18,
A-1050 Vienna, Austria, and\\ D.~V.~Skobeltsyn Institute of
Nuclear Physics, M.~V.~Lomonosov Moscow State University, 119991,
Moscow, Russia, and\\ Faculty of Physics, University of Vienna,
Boltzmanngasse 5, A-1090 Vienna, Austria\\E-mail:
\email{dmitri\_melikhov@gmx.de}}
\author{Hagop Sazdjian\\Institut de Physique Nucl\'eaire,
CNRS-IN2P3, Universit\'e Paris-Sud, Universit\'e Paris-Saclay,
91406 Orsay Cedex, France\\E-mail: \email{sazdjian@ipno.in2p3.fr}}

\abstract{The analysis of two-ordinary-meson scattering amplitudes
in the limit of a large number, $N_{\rm c},$ of the colour degrees
of freedom of quantum chromodynamics, with suitably decreasing
strong coupling and all quarks transforming according to the gauge
group's fundamental representation, enables us to establish a set
of rigorous consistency conditions for the emergence of a
tetraquark (\emph{i.e.}, a bound state of two quarks and two
antiquarks) as a pole in these amplitudes. For genuinely
flavour-exotic tetraquarks, these constraints require the
existence of two tetraquark states distinguishable by their
preferred couplings to two ordinary mesons, whereas, for
cryptoexotic tetraquarks, our constraints may be satisfied by a
single tetraquark state, which then, however, may mix
with~ordinary mesons. For elucidation of the tetraquark features,
the consideration of the \emph{subleading\/} contributions proves
to be mandatory: for both variants of tetraquarks, their decay
widths fall off like $1/N_{\rm c}^2$ for large~$N_{\rm c}.$}

\FullConference{The European Physical Society Conference on High
Energy Physics\\5--12 July 2017\\Venice, Italy}

\begin{document}\section{Incentive: Implications of the
Large-$N_{\rm c}$ Limit of QCD on Polyquark Bound States}
Tetraquark mesons are exotic bound states of two quarks and two
antiquarks hypothesized~to~be predicted by quantum chromodynamics,
the quantum field theory governing the strong interactions. We
extract information on general features of tetraquarks by
considering four-point Green~functions of bilinear quark currents
$j_{ij}=\bar q_i\,\mathfrak{A}\,q_j$ serving as interpolating
operators of mesons $M_{ij}$ composed of antiquark $\bar q_i$ and
quark $q_j$, $\langle 0|j_{ij}|M_{ij}\rangle=f_{M_{ij}}$, where
$i,j,\ldots=1,2,3,4$ represent the flavour~quantum numbers of the
quarks and the generalized Dirac matrix $\mathfrak{A}$ fits to the
interpolated meson's parity and spin quantum numbers, within that
generalization of QCD given by the limit of a large number $N_{\rm
c}$ of colour degrees of freedom, that is, by a quantum field
theory based on the gauge group ${\rm SU}(N_{\rm c})$, with
fermions transforming according to the $N_{\rm c}$-dimensional
fundamental\footnote{For the sake of simplicity, in particular, in
order to deal with a unique $N_{\rm c}\to\infty$ limit, let us
disregard the other~logical possibility of fermions transforming
according to the $\frac{1}{2}\,N_{\rm c}\,(N_{\rm
c}-1)$-dimensional antisymmetric representation of ${\rm
SU}(N_{\rm c})$.} representation of ${\rm SU}(N_{\rm c})$
\cite{Nc}; the strong fine-structure coupling $\alpha_{\rm
s}\equiv g_{\rm s}^2/(4\pi)$ is assumed to decrease,~for~$N_{\rm
c}\to\infty$,~like $\alpha_{\rm s}\propto1/N_{\rm c}$.

Systematic application of the limit $N_{\rm c}\to\infty$ puts us
in a position to discriminate unambiguously two classes of hadrons
according to their large-$N_{\rm c}$ behaviour: those that survive
the large-$N_{\rm c}$ limit~as stable bound states and hence may
be dubbed as ``ordinary'', and the others, \emph{i.e.}, those that
do not but disappear for $N_{\rm c}\to\infty$ \cite{RLJ}. However,
although tetraquarks can only appear at an $N_{\rm c}$-subleading
order \cite{SC}, in order to enable observability by experiment
their decay widths $\Gamma$ should not grow~with~$N_{\rm c}$
\cite{SW}. It is straightforward to prove that at large $N_{\rm
c}$ the meson decay constants $f_{M_{ij}}$ rise like
$f_{M_{ij}}\propto N_{\rm c}^{1/2}$~\cite{Nc}.

\section{Analysis of Tetraquark Poles in Meson--Meson Scattering
Amplitudes at Large $N_{\rm c}$}For well-definiteness, let's base
this large-$N_{\rm c}$ QCD study on a variety of plausible
assumptions:\begin{itemize}\item For the analysis of tetraquarks,
the large-$N_{\rm c}$ limit makes sense and works well; the
application of the $1/N_{\rm c}$ expansion to tetraquarks is
justified and allows us to arrive at reliable conclusions.\item In
the large-$N_{\rm c}$ limit, poles interpretable as tetraquark
bound states exist in the complex plane.\item In the series
expansions in powers of $1/N_{\rm c}$ of those $n$-point Green
functions which potentially accommodate tetraquark poles, the
tetraquark states $T$ arise at the lowest possible $1/N_{\rm c}$
order.\item The masses, $m_T$, of the tetraquark states, $T$, do
not grow with $N_{\rm c}$ but remain finite for $N_{\rm
c}\to\infty$.\end{itemize}

Before embarking on elucidating the dynamics of the formation of a
tetraquark bound state, the main issue is to single out, in the
expansion of a four-point Green function in powers of $1/N_{\rm
c}$ and $\alpha_{\rm s}$, those (large-$N_{\rm c}$) Feynman
diagrams that might develop tetraquark poles. To this end, we
impose, for tetraquarks supposedly consisting of (anti-)quarks of
masses $m_1$, $m_2$, $m_3$ and $m_4$ and, with~respect to the $s$
channel, incoming external momenta $p_1$ and $p_2$ the following
set of basic~selection criteria \cite{TQ}:\begin{enumerate}\item A
tetraquark-phile Feynman diagram depends \emph{nonpolynomially\/}
on its variable $s\equiv(p_1+p_2)^2$.\item A tetraquark-phile
Feynman diagram supports appropriate four-quark intermediate
states and exhibits the corresponding branch cuts starting at the
branch points $s=(m_1+m_2+m_3+m_4)^2$.\end{enumerate} Only Feynman
diagrams complying with both criteria can contribute to the
physical tetraquark~pole.\pagebreak

Under these premises, we derive, for various classes of
tetraquarks $T$, the large-$N_{\rm c}$ behaviour of \textbullet\
the tetraquark-phile four-point Green functions, identified by a
subscript $T$, \textbullet\ the amplitudes $A$ for transitions
between tetraquark $T$ and two ordinary mesons, and \textbullet\
the tetraquark decay rate~$\Gamma(T)$ \cite{TQ}:\begin{itemize}
\item[$\bigstar$] For genuinely \emph{exotic\/} tetraquarks
$T=(\bar q_1\,q_2\,\bar q_3\,q_4)$, involving four different quark
flavours, the correlators without (Fig.~\ref{F:ED}) and with
(Fig.~\ref{F:ER}) a flavour reshuffle behave differently
at~large~$N_{\rm c}$:\begin{eqnarray*}\langle j^\dag_{12}\,
j^\dag_{34}\,j_{12}\,j_{34}\rangle_T=O(N_c^0)\ ,\quad\langle
j^\dag_{14}\,j^\dag_{32}\,j_{14}\,j_{32}\rangle_T=O(N_c^0)\
,&\quad&\langle j^\dag_{14}\,j^\dag_{32}\,j_{12}\,j_{34}\rangle_T
=O(N_c^{-1})\ .\end{eqnarray*}This observation forces us to
conclude that there exist, at least, two different tetraquark
states, called $T_A$ and $T_B$, each with a preferred two-meson
decay channel, but with \emph{parametrically\/}~the same decay
rate of order $N_{\rm c}^{-2}$. Phrased in other words, ``always
two there are, \dots\ no less'' \cite{Y}:\begin{eqnarray*}
A(T_A\leftrightarrow M_{12}\,M_{34})=O(N_{\rm c}^{-1})\ ,\quad
A(T_A\leftrightarrow M_{14}\,M_{32})=O(N_{\rm
c}^{-2})\quad&\Longrightarrow& \quad\Gamma(T_A)=O(N_{\rm c}^{-2})\
,\\A(T_B\leftrightarrow M_{12}\,M_{34})=O(N_{\rm c}^{-2})\ ,\quad
A(T_B\leftrightarrow M_{14}\,M_{32})=O(N_{\rm
c}^{-1})\quad&\Longrightarrow&\quad \Gamma(T_B)=O(N_{\rm c}^{-2})\
.\end{eqnarray*}The tetraquarks $T_A$ and $T_B$ may mix, with
mixing parameter decreasing at least as fast as~$1/N_{\rm c}$.

\begin{figure}[t]\begin{center}\includegraphics[scale=.3521]
{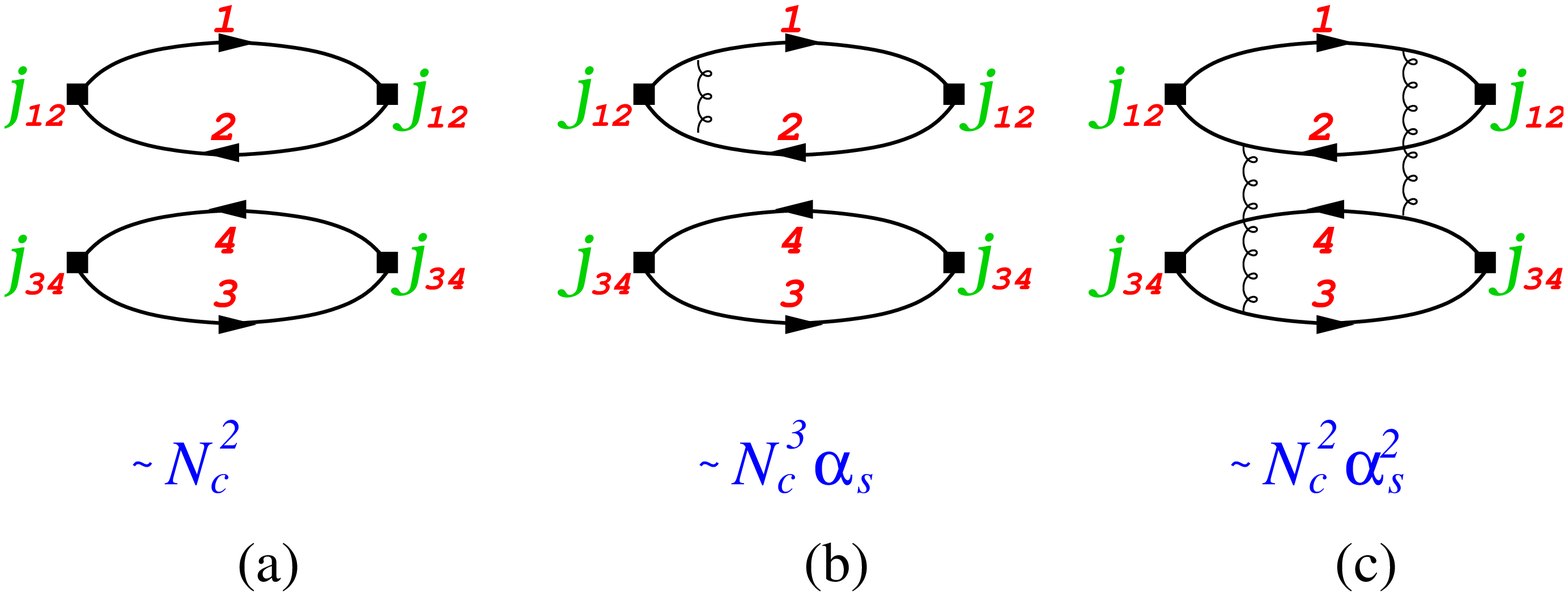}\caption{Four-current Green function
$\langle j^\dag_{12}\,j^\dag_{34}\,j_{12}\,j_{34}\rangle$: $N_{\rm
c}$-leading (a,b) and $N_{\rm c}$-subleading (c) contributions.}
\label{F:ED}\end{center}\end{figure}
\begin{figure}[t]\begin{center}\includegraphics[scale=.3521]
{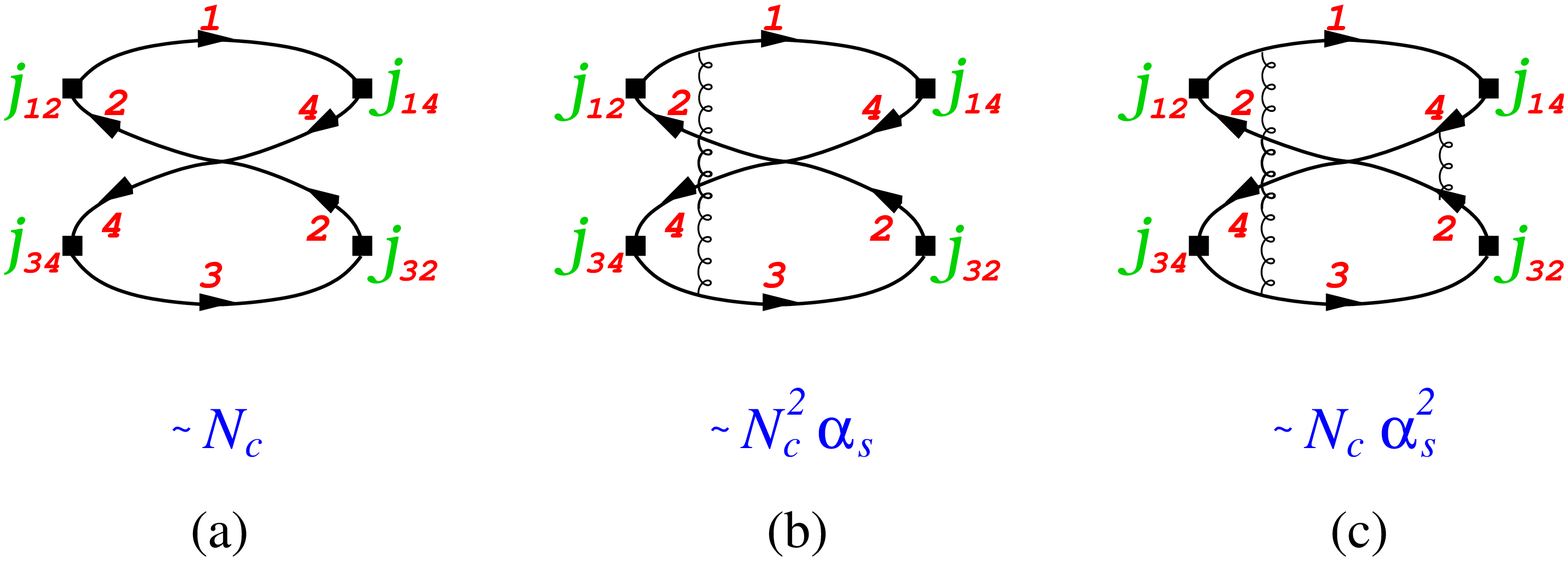}\caption{Four-current Green function
$\langle j^\dag_{14}\,j^\dag_{32}\,j_{12}\,j_{34}\rangle$: $N_{\rm
c}$-leading (a,b) and $N_{\rm c}$-subleading (c) contributions.}
\label{F:ER}\end{center}\end{figure}

\item[$\bigstar$] For \emph{cryptoexotic\/} tetraquarks $T=(\bar
q_1\,q_2\,\bar q_2\,q_3)$, with quark flavour of the ordinary
mesons~$M_{13}$, the correlators without (Fig.~\ref{F:CD}) and
with (Fig.~\ref{F:CR}) a flavour reshuffle have similar $N_{\rm
c}$ behaviour:\begin{eqnarray*}\langle j^\dag_{12}\,j^\dag_{23}\,
j_{12}\,j_{23}\rangle_T=O(N_c^0)\ ,\quad\langle j^\dag_{13}\,
j^\dag_{22}\,j_{13}\,j_{22}\rangle_T=O(N_c^0)\ ,&\quad&\langle
j^\dag_{13}\,j^\dag_{22}\,j_{12}\,j_{23}\rangle_T=O(N_c^0)\ .
\end{eqnarray*}The implied $N_{\rm c}$ constraints may be solved by
a single tetraquark state $T$ decaying according~to
\begin{eqnarray*} A(T\leftrightarrow M_{12}\,M_{23})=O(N_{\rm
c}^{-1})\ ,\quad A(T\leftrightarrow M_{13}\,M_{22})=O(N_{\rm
c}^{-1})\quad&\Longrightarrow& \quad\Gamma(T)=O(N_{\rm c}^{-2})\
,\end{eqnarray*}and mixing with ordinary mesons $M_{13}$ with
mixing strength dropping not slower than $1/\sqrt{N_{\rm
c}}$.\pagebreak

\begin{figure}[t]\begin{center}\includegraphics[scale=.3521]
{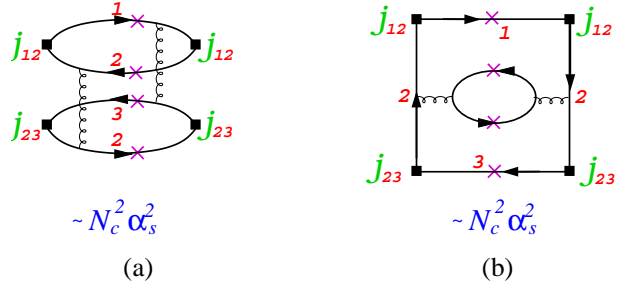}\caption{Four-current Green function
$\langle j^\dag_{12}\,j^\dag_{23}\,j_{12}\,j_{23}\rangle$: some
$N_{\rm c}$-leading contributions potentially capable of
developing a cryptoexotic tetraquark pole, with constituents
identified by purple crosses on their propagators.}\label{F:CD}
\end{center}\end{figure}
\begin{figure}[t]\begin{center}\includegraphics[scale=.3521]
{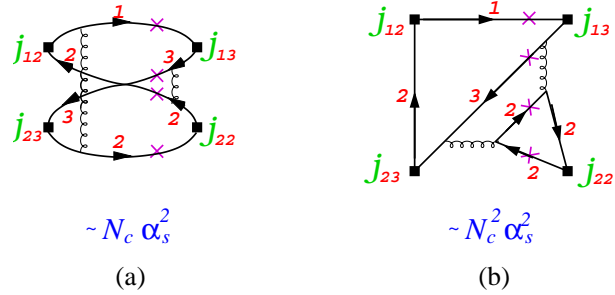} \caption{Four-point Green function
$\langle j^\dag_{13}\,j^\dag_{22}\,j_{12}\,j_{23}\rangle$: $N_{\rm
c}$-leading (b) and $N_{\rm c}$-subleading (a) contributions that
potentially support a cryptoexotic tetraquark pole with quark
content fixed by the purple-crossed propagators.}\label{F:CR}
\end{center}\end{figure}\end{itemize}

\section{Summary: Insights on Minimum Numbers and Decay Rates of
Tetraquark Types}(Crypto-) exotic tetraquarks $T$ \emph{are\/}
narrow: their decay widths $\Gamma(T)$ vanish in the limit $N_{\rm
c}\to\infty$. Unlike earlier claims \cite{MPR}, they have widths
of order $1/N_{\rm c}^2$. If exotic, they come in two versions,
with $N_{\rm c}$-dependent branching ratios. Our results \cite{TQ}
generalize ones got for special cases or channels \cite{EC}.

\noindent\emph{Acknowledgement.} D.M.~was supported by the
Austrian Science Fund (FWF), project P29028-N27.

\end{document}